\input{aipcheck}

\documentclass[
    ,final            
  ]
  {aipproc}

\layoutstyle{6x9}

\begin{document}

\title{Baryon-Strangeness Correlations from Hadron/String- and Quark-Dynamics}

\classification{<\texttt{http://www.aip..org/pacs/index.html}>}

\author{Stephane Haussler}{
  address={Frankfurt Institute for Advanced Studies (FIAS),
Johann Wolfgang Goethe Universit\"at, Max-von-Laue-Str. 1,
60438 Frankfurt am Main, Germany\\}
}

\author{Stefan Scherer}{
  address={Frankfurt Institute for Advanced Studies (FIAS),
Johann Wolfgang Goethe Universit\"at, Max-von-Laue-Str. 1,
60438 Frankfurt am Main, Germany\\}
}

\author{Marcus Bleicher}{
  address={Institut f\"ur Theoretische Physik, Johann Wolfgang Goethe
Universit\"at, Max-von-Laue-Str. 1, 60438 Frankfurt am Main, Germany }
}


\begin{abstract}
Baryon-strangeness correlations ($C_{BS}$) are studied with a hadron/string transport approach (UrQMD) 
and a dynamical quark recombination model (quark molecular dynamics, qMD) for various 
energies from $E_{lab}=4A$~GeV to $\sqrt{s_{NN}}=200$~GeV. 
As expected, we find that the hadron/string dynamics shows correlations similar to a simple hadron gas.
In case of the quark molecular dynamics, we find that initially the $C_{BS}$ correlation is that of a 
weakly interacting QGP but changes in the process of hadronization also to the value for a hadron gas.
Therefore, we conclude that the hadronization process itself makes the initial baryon strangeness correlation 
unobservable. To make an experimental study of this observable more feasible, we also investigate how a restriction to
only charged kaons and $\Lambda$'s (instead of all baryons and all strange particles) influences the theoretical 
result on $C_{BS}$. We find that a good approximation of the full result can 
be obtained in this limit in the present simulation.

\end{abstract}

\maketitle


A plasma of quarks and gluons is believed to be created in the course of the collision of two heavy nuclei 
travelling at ultra-relativistic speeds. Probes based on fluctuations have been proposed throughout the last 
decade to study the properties of QCD-matter close to the phase transition from hadronic to quark degrees of freedom \cite{Stodolsky:1995ds,Shuryak:1997yj,Bleicher:1998wd,Stephanov:1998dy,Mrowczynski:1999sf,Capella:1999uc,Mrowczynski:1999un,Asakawa:2000wh,Muller:2001wj,Cunqueiro:2005hx}. 
Even though they promised to be most adequate due to the strongly fluctuating energy density, initial temperature, 
isospin or particles density  no experimental data up to now 
relying on event-by-event analyses could show a decisive signal for the production of quark-gluon matter (QGP).

A novel event-by-event observable has been introduced by Koch et al. \cite{Koch:2005vg},
the baryon-strangeness correlation coefficient $C_{BS}$. This correlation is proposed as a tool to 
specify the nature (ideal QGP or strongly coupled QGP or hadronic matter) of the highly compressed 
and heated matter created in heavy ions collisions. The idea is that depending on the phase the system 
is in, the relation between baryon number and strangeness will be different: On the one hand, if one 
considers an ideal plasma of quarks and gluons,
strangeness will be carried by freely moving strange and anti-strange quarks, carrying  baryon number 
in strict proportions. This leads to a strong correlation between baryon number and strangeness. On the 
other hand, if the degrees of freedom are of hadronic nature, this correlation is different, because it 
is possible to carry strangeness without baryon number, e.g. in mesons or QGP bound states.

To quantify to which degree strangeness and baryon number are correlated,
the following correlation coefficient has been proposed \cite{Koch:2005vg}:

\begin{equation}\label{eq:definition}
C_{BS}=-3 \frac{\langle BS\rangle-\langle B\rangle\langle S\rangle}{\langle S^{2}\rangle-\langle S\rangle^{2}}\quad,
\end{equation}
where $B$ is the baryon charge and $S$ is the strangeness in a given event. If a QGP is created, the 
value of $C_{BS}$ will be unity as expected from lattice QCD, compatible with the ideal weakly coupled 
QGP. For a hadron gas, where the correlation is non trivial, this quantity has been evaluated 
in  \cite{Koch:2005vg} to be $C_{BS}=0.66$.

In this paper, we study the correlation coefficient $C_{BS}$ with the Ultra-relativistic 
Quantum Molecular Dynamics model (UrQMD v2.2) and the quark Molecular Dynamics model (qMD). 
The UrQMD is a non-equilibrium microscopic transport model that simulates the full space-time 
evolution of heavy ions collisions. It is valid from a few hundreds of MeV to several  TeV per nucleon 
in the laboratory frame. It describes the rescattering of incoming and produced particles, the 
excitation and fragmentation of color strings and the formation and decay of resonances. This model 
has been used before to study event-by-event fluctuations rather successfully \cite{Bleicher:1998wd,Bleicher:1998wu,Bleicher:2000ek,Bleicher:2000tr,Jeon:2005kj,Haussler:2005ei} and yields a reasonable description of inclusive particle 
distributions. For a complete review of the model, the reader is referred to \cite{Bass:1998ca,Bleicher:1999xi}. 
Since the UrQMD is based on hadrons and strings it provides an estimate of the $C_{BS}$ value in the case 
where no QGP is created, however taking into account the rescattering and the non-equilibrium nature of 
the heavy ion reactions.

In contrast, the qMD model provides an out-of-equilibrium estimate of $C_{BS}$ with an 
explicit phase transition from QGP to hadronic matter. It describes the dynamics and the hadronization through an 
effective heavy quark potential in which the quarks propagate with a final dynamical recombination to white clusters. 
These clusters are then mapped to known hadrons and resonances that are later allowed to decay. 
Note that qMD is a recombination model that does not violate energy and momentum conservation and does not 
reduce the entropy in the hadronization process. The reader is referred to \cite{Hofmann:1999jx,Scherer:2001ap} 
for more details about the qMD model.
\begin{figure}
  \includegraphics[height=.35\textheight]{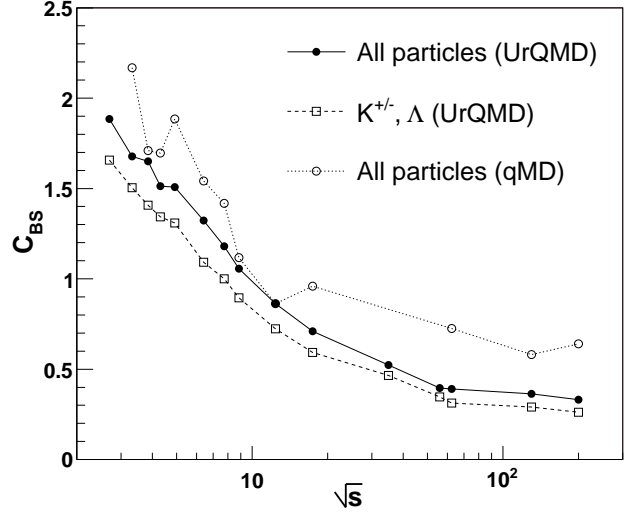}
  \caption{Correlation coefficient $C_{BS}$ for central Au+Au/Pb+Pb as a function of $\sqrt{s}$ calculated with 
UrQMD with all particles taken into account (full circles) and only $\Lambda$'s and charged kaons (open squares). 
Open circles are calculated  with the qMD model using all particles. The rapidity window 
is $y_{\rm max}=0.25$.
\label{fig::excitation}}
\end{figure}

$C_{BS}$ is evaluated from the  event-by-event fluctuation analyses following \cite{Koch:2005vg}:
\begin{equation}\label{eq:calcul}
C_{BS}=-3 \frac{ \frac{1}{N} \sum_{n} B^{(n)} S^{(n)} - (\frac{1}{N} \sum_{n} B^{(n)}) (\frac{1}{N} \sum_{n} S^{(n)}) }
    {\frac{1}{N} \sum_{n} (S^{(n)})^2 - (\frac{1}{N} \sum_{n} S^{(n)})^2}
\end{equation}
$B^{(n)}$ and $S^{(n)}$ stand for the baryon number and strangeness in a given event $n$.

If a QGP is created, the signal given by the $C_{BS}$ coefficient should survive the hadronic phase 
only if the flow is strong enough. I.e. strangeness and baryon number within a given rapidity range 
should be frozen in. 
The rapidity window used must not be too wide in order to avoid global baryon number 
and strangeness conservation which will lead to a vanishing correlation. 
Nevertheless, the acceptance window must be wide enough to avoid smearing 
due to hadronization. A suggested reasonable width is of the order of $y_{\rm max}=0.25-0.5$.

The energy scan of $C_{BS}$ for central Au+Au/Pb+Pb collisions as calculated with UrQMD is shown as full 
circles in Figure \ref{fig::excitation}. As discussed in \cite{Koch:2005vg}, $C_{BS}$ 
increases with an increase of the baryon chemical potential $\mu_{B}$, i.e. when going to lower beam energies. 
With increasing collision energy, and therefore decreasing $\mu_{B}$, $C_{BS}$ goes down to $C_{BS} \approx 0.4$ 
at the highest RHIC energy available and is slightly lower than the value for a fully thermalized hadron gas.
Unfortunately it is difficult to explore $C_{BS}$ directly in experiment, because it includes contributions from
neutrons and other difficult to measure hadrons.
It is therefore desirable to test, if also a better accessible subset of particles can be used to explore
this correlation. Therefore, we study next, how $C_{BS}$ is modified if  only charged kaons 
and $\Lambda$'s are taken into account. As shown in Fig. \ref{fig::excitation} (open squares) one observes that this subset of particles leads in good approximation to the same results for $C_{BS}$ at high energies as for the full
set of hadrons. Therefore, we conclude that a measurement of  the energy dependence 
of the $C_{BS}$ correlation extracted out of charged kaons and $\Lambda$'s only  might 
be sufficient to measure the correlation between baryon number and strangeness in heavy-ions collisions.

Let us finally discuss how a model with quark degrees of freedom compares to the hadron/string dynamics results. 
As shown in Fig. \ref{fig::excitation} (open circles) the result from the quark molecular dynamics model 
follows roughly the shape of the UrQMD values. Especially towards the highest RHIC energy 
$C_{BS}$ decreases below the QGP expectation of $C_{BS}=1$. This surprising result 
strongly contrasts with the expected value for an ideal quark-gluon-plasma and might be a first indication 
that smearing due to hadronization process itself might have drastic effects on fluctuation observables.
\begin{figure}
  \includegraphics[height=.35\textheight]{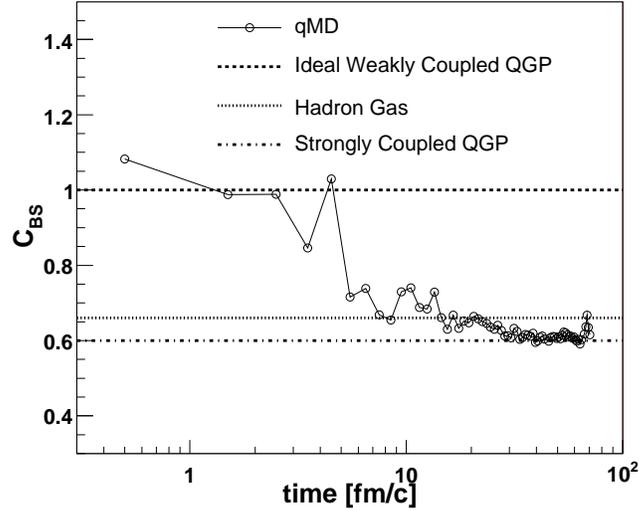}
  \caption{Correlation coefficient from the quark molecular dynamics calculation for Au+Au 
collisions at $\sqrt{s_{NN}}=200$~GeV as a function of 
time at midrapidity. The maximum rapidity 
accepted is $y_{max}=0.25$. Also shown here are the values for an ideal weakly 
coupled QGP, a strongly coupled QGP and for a hadron gas.
    \label{fig::qMD}}
\end{figure}

Let us explore this important question further by studying the time evolution of $C_{BS}$  within the qMD for
Au+Au collisions at $\sqrt{s_{NN}}=200$ GeV as shown in Fig. \ref{fig::qMD}.
At early times one observes that $C_{BS} \approx 1$, in agreement with the expectations for a quark-gluon-plasma. 
However, around 6~fm/c (when the hadronization starts) $C_{BS}$  decreases strongly and reaches its final value
$C_{BS} \approx 0.6-0.7$. One should note that there is no hadronic rescattering stage in qMD, thus,
the decrease of the correlation is solely related to the recombination-like hadronization process in the model.
Thus, the present investigation might explain why no signal of the phase transition has been observed 
in the data up to now. Because even if the initial state consists of a quark-gluon-plasma with the 
expected fluctuations, these fluctuations might be completely blurred in the hadronization process.

To summarize, we have studied the dependence of the baryon-strangeness correlation coefficient
as a function of energy from $E_{lab}=4A$~GeV to $\sqrt{s_{NN}}=200$~GeV for central Au+Au/Pb+Pb 
reactions with two different models. The UrQMD model is based on string-hadronic degrees of freedom, whereas 
the qMD model contains an explicit quark phase and a transition from quark to hadronic matter. 
$C_{BS}$ is found to decrease from the lower energies towards the top RHIC energy in both approaches. 
At the highest RHIC energy  the $C_{BS}$ value from the hadron/string transport model is roughly half 
the one expected in the case of a QGP. 
However, the calculation including a phase transition gives similar results as without phase transition, 
in clear contradiction with what has been expected in case a plasma 
of quarks and gluons as the initial matter. This finding is traced back to the 
hadronization process itself that destroys
the initially present correlations. 

\section*{Acknowledgments}
This work has been supported by GSI and BMBF. The computational resources were provided by the 
Center for Scientific Computing (CSC) in Frankfurt.


\begin{thebibliography}{99}

\bibitem{Stodolsky:1995ds}
  L.~Stodolsky,
  Phys.\ Rev.\ Lett.\  {\bf 75} (1995) 1044.

\bibitem{Shuryak:1997yj}
  E.~V.~Shuryak,
  Phys.\ Lett.\ B {\bf 423} (1998) 9
  [arXiv:hep-ph/9704456].

\bibitem{Bleicher:1998wd}
  M.~Bleicher {\it et al.},
  Nucl.\ Phys.\ A {\bf 638} (1998) 391.

\bibitem{Stephanov:1998dy}
  M.~A.~Stephanov, K.~Rajagopal and E.~V.~Shuryak,
  Phys.\ Rev.\ Lett.\  {\bf 81} (1998) 4816
  [arXiv:hep-ph/9806219].

\bibitem{Mrowczynski:1999sf}
  S.~Mrowczynski,
  Phys.\ Lett.\ B {\bf 459} (1999) 13
  [arXiv:nucl-th/9901078].

\bibitem{Capella:1999uc}
  A.~Capella, E.~G.~Ferreiro and A.~B.~Kaidalov,
  Eur.\ Phys.\ J.\ C {\bf 11} (1999) 163
  [arXiv:hep-ph/9903338].

\bibitem{Mrowczynski:1999un}
  S.~Mrowczynski,
  Phys.\ Lett.\ B {\bf 465} (1999) 8
  [arXiv:nucl-th/9905021].

\bibitem{Asakawa:2000wh}
  M.~Asakawa, U.~W.~Heinz and B.~Muller,
  Phys.\ Rev.\ Lett.\  {\bf 85} (2000) 2072
  [arXiv:hep-ph/0003169].

\bibitem{Muller:2001wj}
  B.~Muller,
  Nucl.\ Phys.\ A {\bf 702} (2002) 281
  [arXiv:nucl-th/0111008].

\bibitem{Cunqueiro:2005hx}
  L.~Cunqueiro, E.~G.~Ferreiro, F.~del Moral and C.~Pajares,
  Phys.\ Rev.\ C {\bf 72}, 024907 (2005)
  [arXiv:hep-ph/0505197].

\bibitem{Koch:2005vg}
  V.~Koch, A.~Majumder and J.~Randrup,
  Phys.\ Rev.\ Lett.\  {\bf 95}, 182301 (2005)
  [arXiv:nucl-th/0505052].

\bibitem{Bleicher:1998wu}
  M.~Bleicher {\it et al.},
  Phys.\ Lett.\ B {\bf 435} (1998) 9
  [arXiv:hep-ph/9803345].

\bibitem{Bleicher:2000ek}
  M.~Bleicher, S.~Jeon and V.~Koch,
  Phys.\ Rev.\ C {\bf 62} (2000) 061902
  [arXiv:hep-ph/0006201].

\bibitem{Bleicher:2000tr}
  M.~Bleicher, J.~Randrup, R.~Snellings and X.~N.~Wang,
  Phys.\ Rev.\ C {\bf 62} (2000) 041901
  [arXiv:nucl-th/0006047].

\bibitem{Jeon:2005kj}
  S.~Jeon, L.~Shi and M.~Bleicher,
  Phys.\ Rev.\ C {\bf 73}, 014905 (2006)
  [arXiv:nucl-th/0506025].

\bibitem{Haussler:2005ei}
  S.~Haussler, H.~Stoecker and M.~Bleicher,
  Phys.\ Rev.\ C {\bf 73}, 021901 (2006)
  [arXiv:hep-ph/0507189].

\bibitem{Bass:1998ca}
  S.~A.~Bass {\it et al.},
  Prog.\ Part.\ Nucl.\ Phys.\  {\bf 41} (1998) 225
  [arXiv:nucl-th/9803035].

\bibitem{Bleicher:1999xi}
  M.~Bleicher {\it et al.},
  J.\ Phys.\ G {\bf 25} (1999) 1859
  [arXiv:hep-ph/9909407].

\bibitem{Hofmann:1999jx}
  M.~Hofmann, M.~Bleicher, S.~Scherer, L.~Neise, H.~Stoecker and W.~Greiner,
  Phys.\ Lett.\ B {\bf 478}, 161 (2000)
  [arXiv:nucl-th/9908030].

\bibitem{Scherer:2001ap}
  S.~Scherer, M.~Hofmann, M.~Bleicher, L.~Neise, H.~Stoecker and W.~Greiner,
  New J.\ Phys.\  {\bf 3}, 8 (2001)
  [arXiv:nucl-th/0106036].

\end{thebibliography}
\end{document}